\documentclass[aps,prx,floatfix,twoside,twocolumn,10pt,citeautoscript,superscriptaddress]{revtex4-2}

\usepackage{amsmath,amssymb}
\usepackage{glossaries}
\usepackage{graphicx}
\usepackage[version=4]{mhchem}
\usepackage[per-mode=symbol]{siunitx}
\usepackage[colorlinks=true]{hyperref}
\hypersetup{
    pdffitwindow=false,
    pdfstartview={FitH},
    pdfnewwindow=true,
    colorlinks=true,
    linkcolor=black,
    citecolor=black,
    filecolor=black,
    urlcolor=black,
}

\graphicspath{{figures}}
\hypersetup{pdfauthor={Eric Lindgren, Adam J. Jackson, Erik Fransson, Esmée Berger, Svemir Rudic, Goran Skoro, Rastislav Turanyi, Sanghamitra Mukhopadhyay, and Paul Erhart}}
\hypersetup{pdftitle={Predicting neutron experiments from first principles: A workflow powered by machine learning}}

\setacronymstyle{long-short}
\newacronym{dft}{DFT}{density functional theory}
\newacronym{md}{MD}{molecular dynamics}
\newacronym{mlip}{MLIP}{machine-learned interatomic potential}
\newacronym{ml}{ML}{machine learning}
\newacronym{nep}{NEP}{neuroevolution potential}
\newacronym{rmse}{RMSE}{root mean square error}
\newacronym{ins}{INS}{inelastic neutron scattering}
\newacronym{sns}{SNS}{spallation neutron source}
\newacronym{pimd}{PIMD}{path-integral molecular dynamics}

\DeclareSIUnit\angstrom{\text{\AA}}
\DeclareSIUnit\Da{\text{Da}}
\DeclareSIUnit\atom{\text{atom}}

\begin{document}

\title{Predicting neutron experiments from first principles: \texorpdfstring{\\}{}A workflow powered by machine learning}

\newcommand{\addChalmers}{Chalmers University of Technology, Department of Physics, 41296 Gothenburg, Sweden}

\newcommand{\addchalmersphysics}{Department of Physics, Chalmers University of Technology, SE-41296, Gothenburg, Sweden}
\newcommand{\addchalmerschemistry}{Department of Chemistry and Chemical Engineering, Chalmers University of Technology, SE-41296, Gothenburg, Sweden}
\newcommand{\addisiscomputing}{Scientific Computing Department, STFC Rutherford Appleton Laboratory, Didcot OX11 0QX, UK}
\newcommand{\addisisneutrons}{ISIS Neutron and Muon Source, STFC Rutherford Appleton Laboratory, Didcot OX11 0QX, UK}

\author{Eric Lindgren}
\affiliation{\addchalmersphysics}
\author{Adam J. Jackson}
\affiliation{\addisiscomputing}
\author{Erik Fransson}
\author{Esmée Berger}
\affiliation{\addchalmersphysics}
\author{Goran \v{S}koro}
\author{Svemir Rudi\'{c}}
\affiliation{\addisisneutrons}
\author{Rastislav Turanyi}
\affiliation{\addisiscomputing}
\author{Sanghamitra Mukhopadhyay}
\affiliation{\addisisneutrons}
\author{Paul Erhart}
\email{erhart@chalmers.se}
\affiliation{\addchalmersphysics}

\begin{abstract}
Machine learning has emerged as a powerful tool in materials discovery, enabling the rapid design of novel materials with tailored properties for countless applications, including in the context of energy and sustainability. 
To ensure the reliability of these methods, however, rigorous validation against experimental data is essential.
Scattering techniques---using neutrons, X-rays, or electrons---offer a direct way to probe atomic-scale structure and dynamics, making them ideal for this purpose.
In this work, we describe a computational workflow that bridges machine learning–based simulations with experimental validation.
The workflow combines density functional theory, machine-learned interatomic potentials, molecular dynamics, and autocorrelation function analysis to simulate experimental signatures, with a focus on inelastic neutron scattering.
We demonstrate the approach on three representative systems: crystalline silicon, crystalline benzene, and hydrogenated scandium-doped \ce{BaTiO3}, comparing the simulated spectra to measurements from four different neutron spectrometers.
While our primary focus is inelastic neutron scattering, the workflow is readily extendable to other modalities, including diffraction and quasi-elastic scattering of neutrons, X-rays, and electrons.
The good agreement between simulated and experimental results highlights the potential of this approach for guiding and interpreting experiments, while also pointing out areas for further improvement.
\end{abstract}

\maketitle

\section{Introduction}

Advancements in materials science are pivotal for technological progress, driving innovations in energy storage, electronics, and catalysis.
Computational methodologies, particularly \gls{dft}, have become essential tools in materials discovery by predicting materials properties and guiding experimental efforts \cite{ShaMarHas24, LiBaoJi24, CheNguLee24, FujHyoShi24, HauFisJai10}. 
The integration of \gls{ml} with these computational techniques has further accelerated the discovery of novel materials, e.g., by enabling rapid screening of vast chemical spaces \cite{CheNguLee24, FujHyoShi24, HuaHuaDon24, GuoYanYu21, LooBalXue19, WuKonKak19}. 
\Gls{ml} has also facilitated the development of \glspl{mlip}, which allow accurate and efficient atomic-scale simulations, bridging the gap between empirical potentials and first-principles methods \cite{MusSukPas23, UnkChmSau21, Beh21}.

However, the predictive power of these computational approaches necessitates rigorous experimental validation.
Scattering experiments, such as neutron, X-ray, and electron scattering, provide critical insights into the structure and dynamics of materials but require precise simulations to interpret the data accurately \cite{CheAndDru21}.
Bridging the gap between computational predictions and experimental observations remains a significant challenge in the field.
Predictive simulations could also significantly enhance experimental planning and execution by ensuring that data acquisition is optimized for maximum information gain while reducing the likelihood of inconclusive or ambiguous results \cite{EhlCroDia22, BorTerBil20, CheKolRam20}.
Furthermore, such simulations can support the preparation of beamline proposals, providing quantitative justifications for instrument time requests by demonstrating expected signal strengths and resolving power.
As experimental facilities increasingly integrate computational tools into their workflows, predictive capabilities are poised to play a crucial role in streamlining the experimental process, improving the overall efficiency of materials characterization, and ultimately accelerating scientific discoveries.

In response to these challenges, we here describe a comprehensive workflow that integrates \gls{dft} calculations, \glspl{mlip} in the \gls{nep} format, \gls{md} simulations using \textsc{gpumd}, \cite{FanWanYin22} the computation of autocorrelation functions via \textsc{dynasor} \cite{FraSlaErh21, BerFraEri25}, and their convolution with atomic form factors, instrument resolution functions and kinematic constraints.
This enables instrument-specific predictions of scattering data from first-principles, allowing direct comparisons between simulations and experimental measurements.

We demonstrate the efficacy of this workflow by applying it to three example systems, including elemental Si, crystalline benzene, and hydrogenated Sc-doped \ce{BaTiO3}, showcasing both its potential for guiding experimental design and accelerating the discovery of new materials as well as its current limitation.
We focus specifically on simulating \gls{ins} experiments, but the general workflow can be easily used to simulate other experimental modalities, including diffraction as well as quasi-elastic and inelastic scattering of neutrons, X-rays, and electrons.

\begin{figure*}
    \centering
    \includegraphics[width=1.0\linewidth]{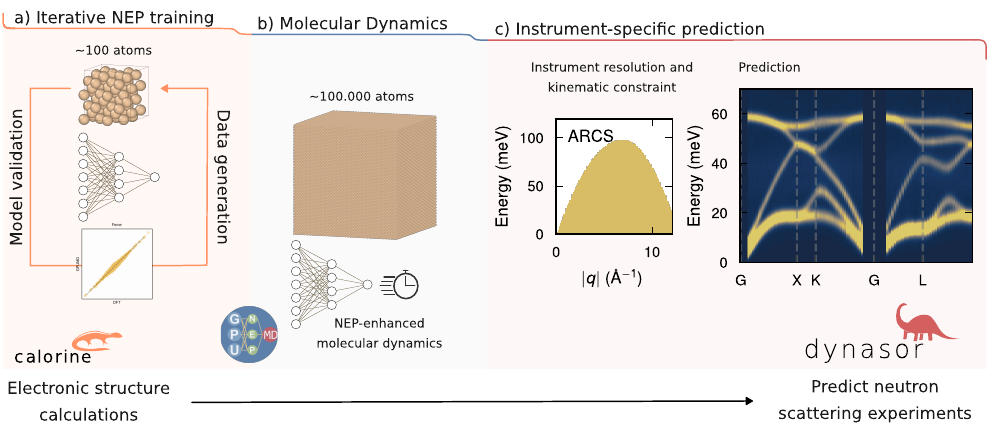}
    \caption{
    Workflow for simulating neutron scattering experiments from first principles. 
    (a) The first step of the workflow comprises constructing \acrfullpl{mlip} using an iterative cycle combining both data generation and model validation.
    Here, the training is facilitated by the \textsc{gpumd} and \textsc{calorine} packages.
    (b) The final \gls{mlip} is used in the second step to run large-scale \gls{md} simulations using the \textsc{gpumd} package.
    (c) The dynamic structure factor is computed from the \gls{md} trajectories using the \textsc{dynasor} package.
    The dynamic structure factor is weighted by species-dependent scattering lengths, and broadened with an instrument-specific resolution function in order to predict the outcome of a particular neutron scattering experiment.
    }
    \label{fig:workflow}
\end{figure*}

\section{Methods}
\label{sect:methodology}

The workflow that we demonstrate for predicting neutron scattering experiments from first principles consists of three steps.
The first step is the the construction of \glspl{mlip} based on the \gls{nep} framework (\autoref{fig:workflow}a; \autoref{sect:mlips}).
These accurate and efficient \glspl{mlip} enable the second step of the workflow, which are large-scale \gls{md} simulations (\autoref{fig:workflow}b; \autoref{sect:md}).
The \gls{md} trajectories that result from the second step are then used in the third step of the workflow, in which we compute the dynamic structure factor that is then weighted by neutron scattering lengths as well as the instrument resolution function and kinematic constraint (\autoref{fig:workflow}c; \autoref{sect:dynasor}).
The weighted dynamic structure factor can then be compared directly to experimental data.

\subsection{Construction of the machine-learned potentials}
\label{sect:mlips}

The workflow starts with training of a \gls{mlip}, or using an already trained model.
Here, we used three different \glspl{mlip} based on the \gls{nep} framework \cite{FanZenZha21, Fan22, FanWanYin22}, one for each of the systems studied in this work. 
For Si, the model published in \cite{FanWanYin22} was used, while we constructed new models for crystalline benzene and hydrogenated Sc-doped \ce{BaTiO3} (\ce{BaTi_{1-x}Sc_xO3H_x}) using the iterative procedure described in Ref.~\citenum{FraWikErh23a} utilizing the \textsc{gpumd} \cite{FanWanYin22} and \textsc{calorine} packages \cite{LinRahFra24} (\autoref{fig:workflow}a).

The training set was initially composed of strained and scaled structures, based on ideal structures using reference data from \gls{dft} calculations (\autoref{sect:dft}).
In the case of benzene, the initial dataset also included dimer configurations to ensure that intermolecular interactions are correctly captured.
An initial model was trained on all available data.
The dataset was then augmented with structures from several iterations of active learning.
To this end, we trained an ensemble of five models by randomly splitting the data into training and validation sets, which was subsequently used to estimate the model uncertainty. 
\gls{md} simulations were then carried out between \num{10} and \qty{200}{\kelvin} and at pressures ranging from \num{0} to \qty{10}{\giga\pascal} for benzene, and from \num{300} to \qty{2000}{\kelvin} and at pressures ranging from \num{-1} to \qty{10}{\giga\pascal} for Sc-doped \ce{BaTiO3} using the respective current generation of \gls{nep} models.
The ensemble was used to select structures with a high prediction uncertainty, quantified by a range of predictions over the ensemble, for which we computed reference energies, forces, and stresses via \gls{dft}.
These configurations were subsequently included when training the next-generation \gls{nep} model.
In total the training set for benzene consisted of \num{798} structures, corresponding to a total of \num{94470} atoms.
For Sc-doped \ce{BaTiO3}, a total of \num{2280} structures, corresponding to a total of \num{138438} atoms.
Structures were generated and manipulated using the \textsc{ase} \cite{LarMorBlo17} and \textsc{hiphive} packages \cite{EriFraErh19}.

We obtained the final \gls{nep} models after \num{13} iterations for benzene and \num{6} iterations for Sc-doped \ce{BaTiO3}.
The final models were trained on all available data.
For the benzene model, the resulting average \glspl{rmse} over the ensemble are \qty{1.1+-0.8}{\milli\electronvolt\per\atom} for the energies and \qty{63+-30}{\milli\electronvolt\per\angstrom} for the forces, and \qty{8.4+-1.6}{\milli\electronvolt\per\atom} for the virials.
The corresponding average coefficients of determination on the same folds are $R^2=\num{0.9997+-0.0003}$, $R^2=\num{0.9949+-0.0051}$, and $R^2=\num{0.9988+-0.0006}$ for energies, forces, and virials, respectively.
The \glspl{rmse} and $R^2$ scores for the final benzene model were \qty{8.510}{\milli\electronvolt\per\atom} and $R^2=\num{0.9998}$ for the energies, \qty{59}{\milli\electronvolt\per\angstrom} and $R^2=\num{0.9962}$ for the forces, and \qty{9.3}{\milli\electronvolt\per\atom} and $R^2=\num{0.9987}$ for the virials (\autoref{si:benzene-training}).
For the Sc-doped \ce{BaTiO3} model, the ensemble \glspl{rmse} were \qty{6.6+-1.3}{\milli\electronvolt\per\atom} for the energies, \qty{186+-34}{\milli\electronvolt\per\angstrom} for the forces, and \qty{31+-5}{\milli\electronvolt\per\atom} for the virials.
The respective coefficients of correlation ($R^2$) were $R^2=\num{0.99998+-0.00001}$, $R^2=\num{0.9765+-0.0058}$, and $R^2=\num{0.9964+-0.0009}$ for energies, forces, and virials, respectively.
The \glspl{rmse} and $R^2$ scores for the final Sc-doped \ce{BaTiO3} model were \qty{6.2}{\milli\electronvolt\per\atom} and $R^2=\num{0.9999}$ for the energies, \qty{172}{\milli\electronvolt\per\angstrom} and $R^2=\num{0.9792}$ for the forces, and \qty{30}{\milli\electronvolt\per\atom} and $R^2=\num{0.9963}$ for the virials (\autoref{si:BTS-training}).

The resulting \gls{nep} models along with the reference data used for training are available via zenodo as specified in the Data Availability statement.

\subsection{Molecular dynamics}
\label{sect:md}

In the second step of the workflow we perform \gls{md} simulations using the \glspl{mlip} from the first step for large supercells (\autoref{fig:workflow}b).
The resulting \gls{md} trajectories are later used to compute the dynamic structure factor as detailed in the next section.

For Si, a supercell comprising \numproduct{38x38x38} primitive cells for a total of \num{438976} atoms was simulated at \qty{300}{\kelvin}, \qty{900}{\kelvin}, \qty{1200}{\kelvin}, and \qty{1500}{\kelvin}, with equilibration of the system in the NPT ensemble and production for \qty{1}{\nano\second} in the NVE ensemble, with a timestep of \qty{2}{\femto\second}.
The atomic positions were written to file every \qty{14}{\femto\second} in order to accurately resolve the fastest vibrations in the system when computing the dynamic structure factor.

Crystalline benzene was simulated in a supercell containing a total of \num{57024} atoms.
The benzene system was equilibrated in the \gls{pimd} ensemble \cite{ParRah84, YinZhoSve25} to avoid the significant underestimation of the cell volume in the classical NPT ensemble at low temperatures.
The \gls{md} simulations were conducted at \qty{127}{\kelvin} to strike a balance between computational cost and the number of \gls{pimd} beads (see \autoref{si:pimd-testing} and \autoref{si:benzene-temperature} in the Supplementary Information).
Production runs were then performed for \qty{1}{\nano\second} in the NVE ensemble.
A timestep of \qty{0.5}{\femto\second} was used, and the positions were written every \qty{3}{\femto\second}.
Ten independent \gls{md} runs were performed to improve the statistics of the computed dynamic structure factor.

Hydrogenated supercells of Sc-doped \ce{BaTiO3} were constructed for various Sc concentrations in the range \num{16}\% to \qty{70}\% in both the cubic and hexagonal phase. 
The supercell contained $\simeq\num{40000}$ atoms.
Equilibration was performed in the \gls{pimd} ensemble at a temperature of \qty{15}{\kelvin}, and production was carried out for \qty{350}{\pico\second} in the thermostated ring-polymer \gls{md} ensemble \cite{RosCerMan14} with a timestep of \qty{0.5}{\femto\second}.
This approach captures nuclear quantum effects on the frequencies \cite{YinZhoSve25}, but it should be noted that the phonon occupation statistics are still classical.
Both equilibration and production runs used 32 \gls{pimd} beads, for an effective system size of $\simeq\num{1300000}$ atoms, limiting the length of the production run compared to Si and benzene because of the increased computational cost.

\subsection{Auto-correlation functions and instrument-specific kinematic constraints}
\label{sect:dynasor}

The central quantity analyzed in the third step of the workflow is the dynamic structure factor, $S(\boldsymbol{q}, \omega)$.
$S(\boldsymbol{q}, \omega)$ is directly proportional to the intensity measured in scattering experiments, and can be readily extracted from \gls{md} simulations. 
While the procedure has been described in detail in \cite{FraSlaErh21, BerFraEri25}, we briefly summarize it here for completeness.
Let $n(\boldsymbol{r}, t)$ denote the particle density defined as 
\begin{align}
    n(\boldsymbol{r}, t) = \sum_i^N \delta \left( \boldsymbol{r} - \boldsymbol{r}_i (t)\right).
\end{align}
$\boldsymbol{r}_i (t)$ is the position of particle $i$ at time $t$, and $N$ is the total number of particles.
The particle density can now be Fourier transformed in space,
\begin{align}
    n(\boldsymbol{q}, t) = \int_{-\infty}^\infty \sum_i^N \delta \left( \boldsymbol{r} - \boldsymbol{r}_i (t)\right) e^{i \boldsymbol{q} \cdot \boldsymbol{r}} \mathrm{d}\boldsymbol{r} = \sum_i^N e^{i \boldsymbol{q} \cdot \boldsymbol{r}_i (t)},
\end{align}
with the autocorrelation function of $n(\boldsymbol{q}, t)$ yielding the intermediate scattering function $F(\boldsymbol{q}, t)$,
\begin{equation}
    F(\boldsymbol{q}, t)
    = \frac{1}{N}\left\langle n(\boldsymbol{q}, t) n(-\boldsymbol{q}, 0)\right\rangle,
\end{equation}
where the brackets denote an ensemble average.
The intermediate scattering function can then be Fourier transformed in time to yield the dynamic structure factor,
\begin{equation}
    S(\boldsymbol{q}, \omega)
    = \int_{-\infty}^\infty  F(\boldsymbol{q}, t) e^{-i\omega t} \mathrm{d}t.
    \label{eq:Sqw_unweighted}
\end{equation}

The dynamic structure factor in Eq.~\eqref{eq:Sqw_unweighted} can be further generalized for multi-component systems. 
Different atomic nuclei scatter neutrons, X-rays, and electrons with varying intensity, which can be taken into account by weighting the partial dynamic structure factor for species $\alpha$ and $\beta$ accordingly.
In the case of neutrons, the partial dynamic structure factor should be weighted by the scattering lengths, $b_\alpha$ and $b_\beta$,
\begin{align}
    S(\boldsymbol{q}, \omega) = \sum_\alpha \sum_\beta b_\alpha b_\beta S_{\alpha \beta}(\boldsymbol{q}, \omega).
    \label{eq:Sqw}
\end{align}

The dynamic structure factor in Eq.~\eqref{eq:Sqw} was computed from the \gls{md} trajectories using the \textsc{dynasor} package \cite{FraSlaErh21, BerFraEri25} in the third step of the workflow (\autoref{fig:workflow}c).
$\boldsymbol{q}$-points and time lags were selected to match the accessible range of the simulated neutron scattering instruments.
Specifically, for Si a Brillouin zone path was sampled connecting the high-symmetry points $\Gamma$, X, K, and L.
The path was sampled in \num{52} different Brillouin zones, randomly selected from the first zone up to $|\boldsymbol{q}|=$\qty{12}{\per/\angstrom} for a total of \num{6136} $\boldsymbol{q}$-points.
Randomly selected $\boldsymbol{q}$-points up to a magnitude $|\boldsymbol{q}|=$\qty{14}{\per\angstrom} and $|\boldsymbol{q}|=$\qty{18}{\per\angstrom} were sampled for benzene and Sc-doped \ce{BaTiO3}, respectively, yielding \num{2116} and \num{2601} $\boldsymbol{q}$-points, respectively.
Gaussian broadening with a width of \qty{0.01}{\per/\angstrom} was then applied to each $\boldsymbol{q}$-point, followed by averaging over spherical shells in $|\boldsymbol{q}|$ to produce $S(q, \omega)$.

Instrument-specific resolution functions and kinematic constraints were applied to the calculated spectra using the \textsc{euphonic} package \cite{FaiJacVon22} with the resolution functions defined in the \textsc{ResINS} package \cite{TurJacWil25}.
The resolution functions used here are Gaussians with energy-dependent width; the functions for TOSCA and Lagrange are based on implementations in \textsc{AbINS}, and the functions for MAPS and ARCS are based on \textsc{PyChop} \cite{DymParFer18}.
(The instrument functions for both \textsc{AbINS} and \textsc{PyChop} are distributed in \textsc{Mantid} \cite{ArnBilBor14}.)
Note that the true resolution functions are four-dimensional and non-Gaussian, but these 1--D approximations are used routinely in \gls{ins} simulations.
The kinematic constraints have their origin in the instrument geometry and transformation from time-of-flight measurements to $(q, \omega)$ space.
In the simulations they are applied as a mask to data computed directly in the $(q, \omega)$ space.

Finally, a quantum correction factor was applied to all dynamic structure factors, in order to correct for the classical phonon statistics generated by the \gls{md} simulations.
Specifically, we applied the following correction factor based on first-order Stokes-Raman scattering \cite{RosFraOst25, CarGun82},
\begin{align}
    S(q, \omega)_{\textrm{corrected}} = S(q, \omega) \frac{\beta \hbar \omega}{1 - \exp{(-\beta \hbar \omega)}}.
    \label{eq:quantum-correction}
\end{align}

\subsection{Density functional theory calculations}
\label{sect:dft}

To generate reference data for the construction of the \glspl{mlip} (\autoref{sect:mlips}) we performed non-spin polarized \gls{dft} calculations using the projector augmented wave method \cite{Blo94, KreJou99} as implemented in the Vienna ab-initio simulation package \cite{KreHaf93, KreFur96, KreFur96a} with a plane wave energy cutoff of \qty{520}{\electronvolt} using the vdW-DF-cx exchange correlation functional \cite{BerHyl14} for benzene and the r2SCAN functional \cite{FurKapNin20} for Sc-doped \ce{BaTiO3}.
The Brillouin zone was sampled with automatically generated $\Gamma$-centered $\boldsymbol{k}$-point grids with an approximate spacing of \qty{0.25}{\per\angstrom} and the partial occupancies in each orbital were set using Gaussian smearing with a width of \qty{0.1}{\electronvolt}.
The \gls{dft} data are available via zenodo as specified in the Data Availability statement.

\subsection{Inelastic neutron scattering experiments on crystalline benzene}

For validation of the predictions for crystalline benzene, inelastic neutron scattering experiments were performed at the TOSCA neutron spectrometer \cite{ParFerRam14, PinRudPar18} at the ISIS Neutron and Muon Source.
The liquid sample was placed in a \qty{1}{\milli\meter} thick standard flat TOSCA aluminum cell which was then briefly submerged into liquid nitrogen.
As soon as the sample solidified it was quickly transferred into the TOSCA closed cycle refrigerator and allowed to further cool to the cryostat base temperature below \qty{10}{\kelvin}.
The short \gls{ins} measurements (approximately 8 minutes per spectrum, i.e., total exposure of \qty{20}{\micro\ampere\hour}) were performed as part of a cooling run at a rate of \qty{3}{\kelvin\per\minute}, with the initial spectrum taken at a starting temperature of \qty{127}{\kelvin} and followed by other measurements at a starting temperature of \qty{103}{\kelvin}, \qty{75}{\kelvin}, \qty{46}{\kelvin}, and \qty{24}{\kelvin}.
The longer \gls{ins} measurement  (approximately 2 hours, i.e., total exposure of \qty{285}{\micro\ampere\hour}) was performed at the base temperature of \qty{10}{\kelvin}, giving a superior spectral signal-to-noise ratio.
The raw data, i.e., time-of-flight events, were reduced using \textsc{Mantid} \cite{ArnBilBor14}.

\subsection{Post-processing and plotting}

The \gls{nep} models and calculated correlation functions were post-processed and analyzed using \textsc{python} scripts, utilizing the \textsc{numpy} \cite{HarMilvan20}, \textsc{pandas} \cite{Mck10, The24}, and \textsc{scipy} \cite{VirGomOli20} packages.
Plots were generated using \textsc{matplotlib} \cite{Hun07}, with color maps from \textsc{perfect-cmaps} \cite{Ulm25}.
Atomic structures were visualized and analyzed using \textsc{Ovito} \cite{Stu09}.

\section{Results}

\begin{figure*}
    \centering
    \includegraphics[width=1.0\linewidth]{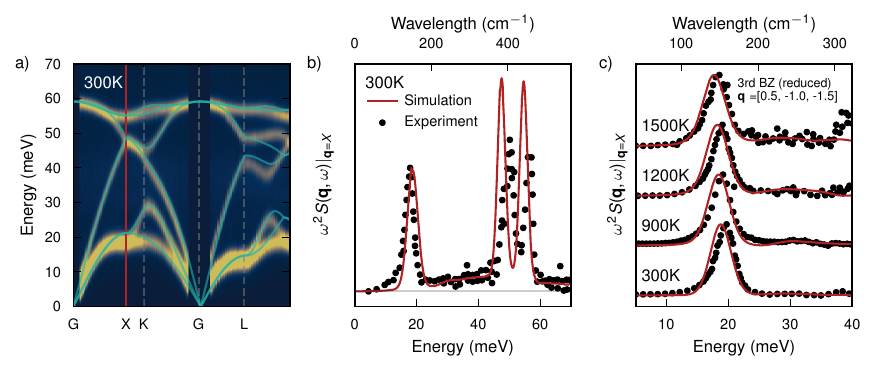}
    \caption{
    (a) Simulated \gls{ins} dispersion of Si from \gls{md} for the ARCS spectrometer at \acrlong{sns}, with the harmonic phonon dispersion calculated using the \gls{nep} model overlaid (turquoise lines).
    (b) The simulated intensity at the X-point, as well as (c) the first peak in the intensity at the X-point in the third Brillouin zone for \qty{300}{\kelvin}, \qty{900}{\kelvin}, \qty{1200}{\kelvin}, and \qty{1500}{\kelvin} (red lines).
    Note that the dispersion in (a) and intensity in (b) is aggregated over 52 different Brillouin zones, while in (c) the results for only a single Brillouin zone are shown.
    The experimental data (black circles) is from \cite{KimHelHer18}.
    The \gls{md} simulation captures both the anharmonicity and multi-phonon effects present in the experimental data, as well as the mode softening as the temperature is increased.
    The multi-phonon effects manifest as non-zero intensity between the acoustic and optical branches at the X-point.
    }
    \label{fig:anharmonicity}
\end{figure*}

\subsection{Anharmonicity in Si}

\begin{figure}
    \centering
    \includegraphics[width=1.0\linewidth]{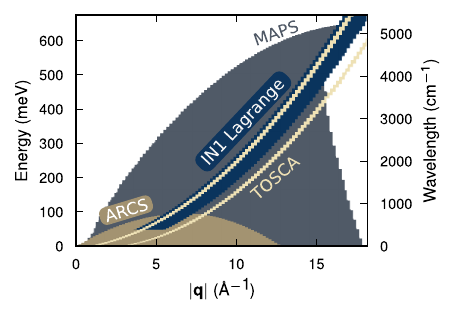}
    \caption{
    Kinematic constraints for the four neutron instruments simulated in this work. 
    ARCS (BL-18) is a wide-range spectrometer at the \acrfull{sns} at the Oak Ridge National Laboratory (USA).
    TOSCA and MAPS are spectrometers at the ISIS Muon and Neutron Source (UK).
    Finally, IN1 Lagrange is a spectrometer at the \emph{Institute Laue-Langevin} in Grenoble (France).
    Note that all spectrometers differ in $q$ and energy range and resolution, owing to their respective kinematic constraints and resolution functions.
    }
    \label{fig:instruments}
\end{figure}

We begin by applying the workflow outlined in the methodology section to simulate an \gls{ins} experiment on crystalline Si at \qty{300}{\kelvin} reported in Ref.~\citenum{KimHelHer18}, that was carried out at the ARCS wide range angular spectrometer (BL-18) at the \acrfull{sns} at Oak Ridge National Laboratory (\autoref{fig:anharmonicity}a).
The simulation is made instrument-specific by applying the resolution function and kinematic constraint for ARCS to the simulated dynamic structure factor.
The $q$ and energy range measured by ARCS is relatively broad (\autoref{fig:instruments}), and thus we computed the dynamic structure in multiple different Brillouin zones to accurately sample the full range allowed by the instrument.
In total, \num{52} Brillouin zones were sampled. 
The dynamic structure factors $S(\boldsymbol{q}, \omega)$ were then weighted by a factor $\propto 1/|\boldsymbol{q}|^2$ since to first order the scattered intensity grows as $|\boldsymbol{q}|^2$.
Furthermore, the Debye-Waller factor, $\exp{(-q^2 U /3)}$, was corrected for in each of the Brillouin zones before the zones were averaged together.
This expression for the Debye-Waller factor assumes an isotropic displacement of the atoms in all Cartesian directions with $U = \left<u^2 \right>$ being the mean squared displacement in the system \cite{Squ12}. 
$U$ was estimated to be \qty{0.013}{\angstrom\squared} from a \qty{100}{\pico\second} \gls{md} simulation of the Si system at \qty{300}{\kelvin}, otherwise following the same protocol as the other simulations of Si in this work.

We validate our results by comparing them to the experimental data measured on the ARCS spectrometer in Ref.~\citenum{KimHelHer18}.
Specifically, we consider the intensity at the high-symmetry X-point (\autoref{fig:anharmonicity}b), where the simulated intensity has been multiplied by an extra factor of $\omega^2$.
Our results are in quantitative agreement with experiments, with the centroids of the phonon mode peaks agreeing well.
The relative intensity between the different phonon mode peaks in the experiment are not entirely reproduced in the simulation, which could be due to a missing correction factor or experimental variability.

We note, in particular, the nonzero intensity measured in the experiment and captured by the simulation in the region \qty{30}{\milli\electronvolt} to \qty{50}{\milli\electronvolt}.
This scattered intensity corresponds to multi-phonon effects, which are inherently captured by \gls{md} simulations.
Furthermore, the effect of thermal expansion as the temperature is varied is also directly included by the \gls{md} simulations, where specifically the low-energy mode at the X-point is softened as temperature is increased (\autoref{fig:anharmonicity}c).
One can observe that some of the predicted mode energies are slightly shifted compared to experiments, by approximately \qty{1}{\milli\electronvolt}.
Given that the \gls{mlip} accurately reproduces the harmonic phonon dispersion from \gls{dft}, it is most likely due to the underlying exchange-correlation functional.

\begin{figure}
    \centering
    \includegraphics[width=1.0\linewidth]{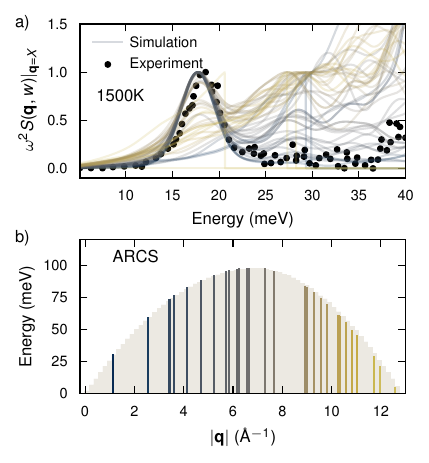}
    \caption{
    (a) Simulated intensity at the X-point at \qty{1500}{\kelvin} in crystalline Si for each of the 52 Brillouin zones accessible by the ARCS spectrometer.
    Note that the intensity of the multi-phonon shoulder at \qty{30}{\milli\electronvolt} varies greatly with the Brillouin zone.
    (b) The kinematic constraint for the ARCS spectrometer at \gls{sns}, which limits the range of non-zero intensities in (a) depending on the $|\boldsymbol{q}|$ for the X-point in each Brillouin zone.
    }
    \label{fig:higher-bz}
\end{figure}

At \qty{1500}{\kelvin} it moreover appears that in the region around \qty{40}{\milli\electronvolt} the simulated and experimental \emph{intensities} differ.
This discrepancy could be due to the Brillouin zone in which the simulations have been conducted.
Here, we show the X-point in the third Brillouin zone, i.e., $\boldsymbol{q}=[0.5, -1.0, -1.5]$, while in the experimental reference \cite{KimHelHer18} the exact $\boldsymbol{q}$-point is not specified.
In fact, the intensity at the X-point varies substantially depending on the Brillouin zone, especially the intensity of the multi-phonon shoulder around \qty{30}{\milli\electronvolt} (\autoref{fig:higher-bz}).
For the comparison shown in (\autoref{fig:anharmonicity}c), we selected the Brillouin zone for which the simulated spectrum best reproduces the experimental data, based on the mean-squared error calculated over the spectrum.

In \gls{md}, the dynamics of the system described by the potential model is captured at the classical level, including high-order phonon effects, thermal expansion, and full anharmonicity.
Efforts have been made in recent years to include the effects of anharmonicity on top of harmonic models, including but not limited to using higher-order force constants \cite{EriFraErh19}, temperature-dependent effective potentials \cite{HelAbr13}, anharmonic lattice models \cite{TadGohTsu14}, and the self-consistent harmonic approximation \cite{MonBiaChe21}. 
However, a harmonic model is inherently limited in describing such complicated dynamic events.

\begin{figure*}
    \centering
    \includegraphics[width=1.0\linewidth]{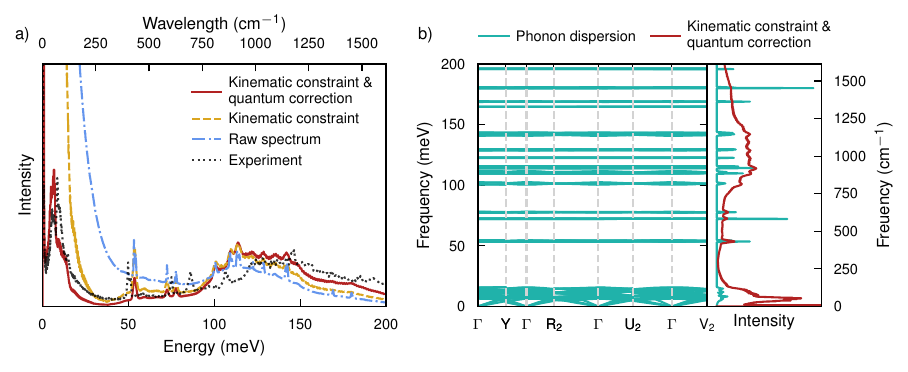}
    \caption{
    (a) Simulated \gls{ins} spectra for crystalline benzene, at increasing levels of refinement, compared to an experimental spectrum obtained at \qty{127}{\kelvin} at the TOSCA spectrometer at the ISIS Neutron and Muon Source (UK).
    The first level of accuracy is the raw simulated spectrum from \gls{md} with only scattering lengths applied (raw spectrum).
    Correcting for the resolution function and kinematic constraint of the TOSCA spectrometer yields a marked increase in accuracy, with further improvement when additionally applying a quantum correction factor to compensate for the classical statistics in \gls{md} simulations, especially for low energies around \qty{10}{\milli\electronvolt}.
    The spectra have been individually scaled to match the experimental spectrum as closely as possible above \qty{50}{\milli\electronvolt}.
    (b) Phonon dispersion and density of states compared to the simulated \gls{ins} spectrum at \qty{127}{\kelvin}, corrected for the kinematic constraint of the TOSCA spectrometer and with quantum statistics.
    }
    \label{fig:corrections}
\end{figure*}

\subsection{Corrections in crystalline benzene}

We now turn to simulating an \gls{ins} measurement of crystalline benzene at \qty{127}{\kelvin} at the TOSCA spectrometer at the ISIS Neutron and Muon Source (UK) in order to study the effects of the resolution function and quantum correction in more detail (\autoref{fig:corrections}a).

The scattered intensity from benzene is dominated by incoherent scattering from hydrogen, owing to the exceptionally large incoherent scattering length of hydrogen.
We thus study the \gls{ins} spectrum directly.
The dynamic structure factor can be integrated over $|\boldsymbol{q}|$ in order to obtain $S(\omega) = \int S(q, \omega) \mathrm{d}q$.
Comparing the raw simulated spectrum with the experimental data, we find that the simulated spectrum captures the peaks corresponding to different modes but the relative intensity between them is not reproduced.
Furthermore, the low-energy peak at \qty{10}{\milli\electronvolt} is not captured. 
The reason for these discrepancies is to a large extent due to the kinematic constraint and resolution function of the TOSCA spectrometer (\autoref{fig:instruments}).
The two detector banks of TOSCA map out two lines in $q$--$\omega$ space, where high (low) frequencies correspond to large (low) $q$.
By sampling along these {$q$--$\omega$} lines and convoluting the resulting spectrum with the resolution function of the instrument, the agreement improves notably.

However, the ratio in intensity between the high and low-energy regions is still not reproduced.
The main reason for this discrepancy is the classical statistics of \gls{md} simulations, which we correct for with the quantum correction factor according to Eq.~\eqref{eq:quantum-correction}.
Applying both kinematic constraint and quantum correction yields a simulated spectrum that is in near-quantitative agreement with experiments. 
The remaining difference to experiments is a redshift of the simulated spectrum by approximately \qty{25}{\milli\electronvolt}. 
We attribute this redshift to the \gls{dft} functional used to train the \gls{nep} model.
A more detailed discussion comparing experiments and first-principles calculations to the predictions from the \gls{nep} model can be found in the Supplementary Information (\autoref{si:benzene-red-shift}).

We can further elucidate the simulated \gls{ins} spectrum by comparing it to the phonon dispersion according to the underlying \gls{mlip} (\autoref{fig:corrections}b).
The simulated \gls{ins} spectrum differs notably from the phonon density of states in terms of intensity, owing to the kinematic constraint of the TOSCA spectrometer, and the quantum correction.
Furthermore, the full anharmonicity included in the \gls{md} simulation in combination with the TOSCA resolution function yields a broadening of the peaks in the simulated \gls{ins} spectrum.

In summary, this study of crystalline benzene highlights the importance of considering the resolution and kinematic constraints of the specific instrument, as well as correcting the statistics from classical \gls{md} simulations, when aiming for quantitative predictions of neutron scattering experiments.

\subsection{Hydrogen dynamics in hydrogenated Sc-doped \texorpdfstring{\ce{BaTiO3}}{BaTiO3}}

\begin{figure*}
    \centering
    \includegraphics[width=1.0\linewidth]{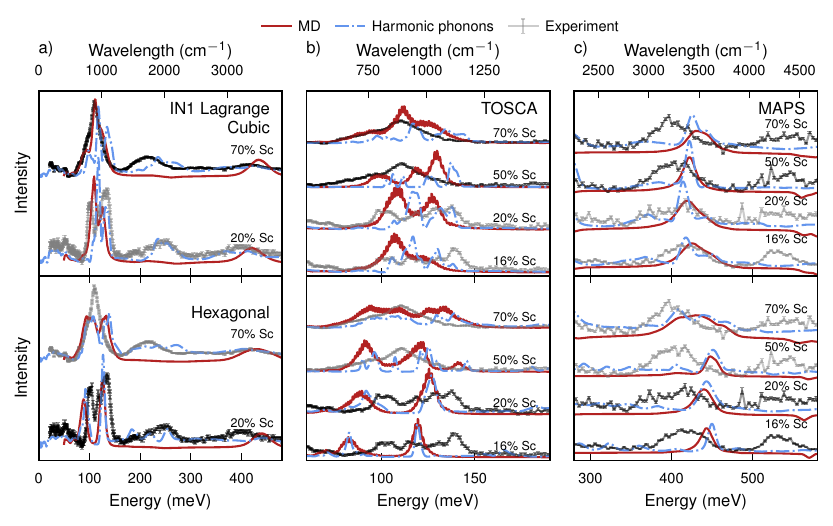}
    \caption{
    Hydrogen dynamics in hydrogenated Sc-doped \ce{BaTiO3} (\ce{BaTi_{1-x}Sc_xO3H_x}) for various concentrations of dopants, compared with experimental results measured at three different neutron spectrometers: IN1 Lagrange, TOSCA, and MAPS.
    Experimental data are from \cite{PerTorJed20}.
    (a) IN1 Lagrange is a wide-range spectrometer that probes the dynamics in the region from \qty{0} to \qty{500}{\milli\electronvolt}.
    (b) TOSCA and (c) MAPS on the other hand can be used to study the region around \qty{100}{\milli\electronvolt} and \qty{500}{\milli\electronvolt}, respectively, utilizing the higher energy resolution they offer.
    Simulated spectra using our workflow (labeled MD) corrected for quantum statistics and the kinematic constraint of the specific instrument are shown for all concentrations of dopants considered experimentally, with the simulated structure both in the cubic and hexagonal structure.
    Experimentally, only one of these phases is stable for a given concentration of Sc, but the energy of the two phases are sufficiently close that both phases are stable on the timescales of the \gls{md} simulations.
    The experimental data has thus been duplicated in the upper and lower rows of plots, where the compositions that are stable for each phase are indicated by the black lines.
    Additionally, simulated spectra based on harmonic phonons obtained via \textsc{AbINS} are included for comparison.
    }
    \label{fig:batio3}
\end{figure*}

Finally, we turn to a more complicated system, in the form of hydrogenated Sc-doped \ce{BaTiO3} (\ce{BaTi_{1-x}Sc_xO3H_x}) where $x$ is the doping fraction of the tetravalent site (Ti, Sc).
Perrichon \textit{et al.} have performed a detailed \gls{ins} study of the hydrogen dynamics in this system at three different spectrometers: the TOSCA and MAPS spectrometers at the ISIS Neutron and Muon source as well as IN1 Lagrange at the \emph{Institut Laue-Langevin} \cite{PerTorJed20}.
The experiments were carried out at temperatures below \qty{20}{\kelvin}.
\Gls{ins} spectra were then obtained by averaging the dynamic structure factor $S(q, \omega)$ up to a magnitude of $|\boldsymbol{q}|=$\qty{12}{\per\angstrom}.
Our simulations presented in this section were averaged over $q$ up to the limit of the kinematic constraint for MAPS, $|\boldsymbol{q}|=$\qty{18}{\per\angstrom}, in order to obtain better statistics.

Sc-doped \ce{BaTiO3} undergoes a phase transition from a hexagonal structure to a cubic perovskite structure as the Sc concentration increases.
On \gls{md} time scales, both structures are, however, at least metastable over the entire composition range, which (in contrast to experiment) allows us to sample structure and composition independently (\autoref{fig:batio3}).

The simulated spectra using our workflow and the experimental spectra agree well in the full energy range \qty{0} to \qty{500}{\milli\electronvolt}. 
The peak at \qty{125}{\milli\electronvolt} corresponds to O--H vibrations, and is best described by the hexagonal phase for low Sc concentrations and by the cubic phase for high Sc-concentrations (\autoref{fig:batio3}a).
However, the overtone peak at \qty{250}{\milli\electronvolt} is underestimated in both simulated phases.
This discrepancy could be due to the quantum correction factor in Eq.~\eqref{eq:quantum-correction}, which is only valid for first-order scattering.
This is further supported by the simulated spectra obtained using \textsc{AbINS}, which accurately capture the intensity of the \qty{250}{\milli\electronvolt} overtone peak.
The latter method handles multi-phonon effects perturbatively and includes quantum effects but does not account for anharmonicity, which explains the sharper first-order features compared to the \gls{md}-based simulations.


The results from TOSCA highlight the \qty{125}{\milli\electronvolt} feature further (\autoref{fig:batio3}b).
For low Sc concentrations the simulated spectrum using our workflow for the hexagonal structure agrees well with experiments, although the simulated spectrum is redshifted by approximately \qty{25}{\milli\electronvolt}.
The simulated spectrum for the cubic structure agrees better with experiments as the Sc  concentration is increased, which is in line with the hexagonal to cubic phase transition with increasing Sc concentration.
We can thus clearly distinguish the spectra for the two phases of Sc-doped \ce{BaTiO3}, as the Sc-doping is varied.

\begin{figure}
    \centering
    \includegraphics[width=1.0\linewidth]{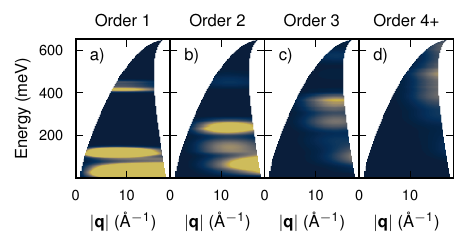}
    \caption{
    Contribution to the total dynamic structure factor from different scattering orders for hydrogenated Sc-doped \ce{BaTiO3} with 16\% Sc in the cubic phase, simulated for the MAPS spectrometer using \textsc{AbINS}. 
    (a) Fundamental modes with higher orders corresponding to overtones and combination modes in panels (b--d).
    The color scale is the same for all subpanels, which leads to clipping of the intensity in panel (a).
    Specifically note that the feature at \qty{550}{meV} originates from fourth-order scattering processes and above.
    }
    \label{fig:abins-orders}
\end{figure}

Finally, the MAPS spectrometer probes the high-energy region between \qty{300}{\milli\electronvolt} to \qty{600}{\milli\electronvolt}.
The feature in the experimental spectra at \qty{450}{\milli\electronvolt} corresponds to stretching of the O--H bond according to Perrichon \emph{et al.}, with the peak at \qty{550}{\milli\electronvolt} assigned as a combination mode of the O--H wag mode at \qty{120}{\milli\electronvolt} and the O--H stretch mode at \qty{450}{\milli\electronvolt}.
The fundamental vibrational peak at \qty{450}{\milli\electronvolt} is captured by the simulations, although with a slight blueshift of \qty{10}{\milli\electronvolt}.
However, the intensity for the combination mode is not reproduced by the simulation, neither using the \gls{md}-based workflow nor \textsc{AbINS}.
In this case, we can further elucidate the nature of the combination modes at \qty{550}{\milli\electronvolt} using \textsc{AbINS} (\autoref{fig:abins-orders}).
In these harmonic incoherent-approximation simulations,
the intensity in that region is mainly composed of fourth-order scattering events and above.
Such high-order phonons require a higher-order correction factor in order for the statistics to come out correctly using the \gls{md}-based workflow.
However, applying a higher-order correction factor is not straightforward, as one would have to know \emph{a priori} in which region of the spectrum to apply the correction, and the order of the higher-order scattering process.

\subsection{Discussion}

The demonstrated workflows enable predictions of neutron scattering experiments, here in the form of \gls{ins} spectra, from first principles.
Starting from an atomic representation of a material, we develop \glspl{mlip} using the \gls{nep} framework for performing accurate \gls{md} simulations for systems comprising at least tens of thousands of atoms over at least a few nanoseconds, from which the dynamic structure factor can be computed using the \textsc{dynasor} package.
We have focused on crystalline materials in the examples above, but the workflow is directly applicable to disordered systems as well, including liquids as well as amorphous or biological materials.
The predictions are made instrument-specific by applying resolution functions and kinematic constraints, and are additionally corrected to account for the classical statistics inherent to the \gls{md} simulations.

The predictions show remarkable quantitative agreement with experiments.
Almost all experimental features in the form of vibrational peaks are faithfully reproduced in the predicted spectra, including their relative intensities after applying correction factors for the quantum statistics.
Remaining differences between the spectra, such as systematic blue- or redshifts of certain peaks in the case of crystalline benzene, can be most likely attributed to the \gls{dft} functional used for training the \gls{mlip}.
However, the intensity for some higher-order phonon processes, such as in Sc-doped \ce{BaTiO3}, are not reproduced faithfully compared to experiments at the present level.
In principle these discrepancies can, however, be corrected for by applying higher-order correction factors to recover the correct statistics for overtones and combination modes.

Such a correction would not represent a general-purpose \emph{ab initio} approach as it is only possible in regions where these features can clearly be identified and separated, such that a single correction can be applied.
For an unknown system, identifying such modes is problematic, although some information can be gained from harmonic calculations, e.g., using the \textsc{AbINS} package.
However, these corrections only affect the relative intensity of these peaks, the positions are directly obtained from the \gls{md} simulations.
Applying higher-order quantum corrections is thus not strictly necessary in order to give a reasonable prediction of a neutron scattering experiment.
In general, we suggest applying the lowest order correction for the whole spectrum to be sufficient for the purpose of guiding neutron scattering experiments.

The workflow as presented in this work relies on \glspl{mlip} in order to run accurate and efficient \gls{md} simulations.
Classical force fields could be used but the results might be of limited accuracy, especially in systems involving both bonded and non-bonded interactions.
However, training a \gls{mlip} or selecting an appropriate force field for a system of interest constitutes a bottleneck in the workflow, requiring domain knowledge and effort.
Foundational models trained on large parts of the periodic table, such as MACE-MP-0 or CHGNet among others \cite{CheOng22, DenZhoJun23, MerBatSch23, XieLuMen24, BatBenChi24}, offer an appealing alternative to creating bespoke \glspl{mlip} or using existing force fields.
These foundational models can either be used out-of-the-box, or fine tuned with a small number of structures from \gls{dft} to yield an accurate model with comparatively low effort.
It should, however, be noted that these models are computationally much more demanding than either \gls{nep} models or classical force fields.

\section{Conclusion}

In this study, we have presented a workflow that enables predictions of neutron scattering experiments from first principles, by combining \gls{dft} calculations, \glspl{mlip}, autocorrelation functions from \gls{md} simulations as well as instrument resolution functions and kinematic constraints.
We envision this workflow to be of great use in the context of materials discovery, offering an avenue for generating simulated experimental signatures for novel materials that can be directly compared to neutron, X-ray, and electron scattering experiments.
\Gls{ml} in the form of \glspl{mlip} plays a central role, as the latter enable the accurate \gls{md} simulations and the extensive sampling that are the foundation of the present workflow.
By integrating these components into a cohesive pipeline, our approach bridges the gap between theory and experiment, facilitating a more efficient feedback loop in the design and characterization of new materials. 
Ultimately, this workflow stands to accelerate materials analysis and discovery processes by providing high-fidelity, simulation-based insights that are directly aligned with experimental observables.

\section{Conflicts of interest}

There are no conflicts to declare.

\section{Data availability}

The \gls{mlip} models, training data and databases of \gls{dft} calculations for benzene and hydrogenated Sc-doped \ce{BaTiO3}, as well as the reduced \gls{ins} data for crystalline benzene are available on zenodo at \url{https://doi.org/10.5281/zenodo.15283533}.
The development of \textsc{gpumd} package is hosted at \url{https://github.com/brucefan1983/GPUMD} and its documentation can be found at \url{https://gpumd.org}.
The \textsc{calorine} package is hosted at \url{https://gitlab.com/materials-modeling/calorine}, its documentation is provided at \url{https://calorine.materialsmodeling.org}, and releases are available at \url{https://doi.org/10.5281/zenodo.7919206}.
The \textsc{dynasor} package is hosted at \url{https://gitlab.com/materials-modeling/dynasor}, its documentation is provided at \url{https://dynasor.materialsmodeling.org}, and releases are available at \url{https://doi.org/10.5281/zenodo.10012241}.

\section{Acknowledgments}

We gratefully acknowledge funding from the Swedish Foundation for Strategic Research via the SwedNESS graduate school (GSn15-0008) and the Swedish Research Council (Nos.~2020-04935 and 2021-05072) as well as computational resources provided by the National Academic Infrastructure for Supercomputing in Sweden at NSC, PDC, and C3SE partially funded by the Swedish Research Council through grant agreement No.~2022-06725, as well as the Berzelius resource provided by the Knut and Alice Wallenberg Foundation at NSC.
We are grateful to the Science and Technology Facilities Council (STFC), and to the ISIS Neutron and Muon Source in particular, for the provision of beamtime via TOSCA Xpress Access route (RB1990290).

\end{document}